\documentclass{aa}  
\usepackage{graphicx}
\usepackage{txfonts}
\usepackage{lipsum}
\usepackage{subcaption}                                        
\usepackage{lscape}                     
\usepackage{placeins}           
\usepackage{CJK}

\makeatletter
\let\do@mlinenumbers\relax
\let\if@linenumbers\iffalse
\let\linenumbers\relax
\let\runningpagewiselinenumbers\relax

\makeatother

\begin{document}
\begin{CJK*}{UTF8}{gbsn}
   \title{Insights into the 3D layered structure of nearby open clusters through N-body simulations}

   \author{Kaixiang Lang (郎凯祥)\inst{1,2}
        \and Yu Zhang (张余)\inst{1,2}
        \and Hubiao Niu (牛虎彪)\inst{1}
        \and Jayanand Maurya\inst{1}\\
        \and Jinzhong Liu(刘进忠)\inst{1}
        \and Guimei Liu (刘桂梅)\inst{1,2}
        }

   \institute{Xinjiang Astronomical Observatory, Chinese Academy of Sciences, 
              No. 150, Science 1 Street, Urumqi, Xinjiang 830011, PR China\\
              \email{zhy@xao.ac.cn}
        \and School of Astronomy and Space Science, University of Chinese Academy of Sciences, 
              19 Yuquan Road, Shijingshan District, Beijing 100049, PR China\\ }

   \date{Received 7 February 2025 / Accepted 6 April 2025}  

  \abstract
   {Open clusters (OCs) are important for understanding star formation, dynamics, and evolution. Previous studies have indicated a relationship between cluster structure and member star properties, but the formation mechanism of the layered structure of OCs remains unclear.}
   {We study the three-dimensional spatial distribution of 279 nearby OCs to understand the formation mechanism of the layered structure.} 
   {We analyzed the spatial distribution of member stars within each OC and correlated the presence of a layered structure with the number of member stars. Additionally, we performd $N$-body simulations to model the evolution of OCSN 125. We assessd the correlation between the binary fraction, the most massive star, and the radius of the layered structure in each simulated OC.}
   {Our analysis reveals that OCs with fewer member stars tend to lack a layered structure. The results from $N$-body simulations indicate that the presence of a layered structure is strongly influenced by dynamical factors, particularly the most massive star and the binary fraction. Massive stars drive mass loss through supernova explosions and stellar winds, which weaken the spatial layering. Furthermore, clusters with higher binary fractions exhibit a weaker layered structure, likely due to energy equipartition, dynamical friction, and perturbations caused by binary systems. These factors contribute to delaying core collapse and slowing the emergence of a layered structure.}
   {Our findings suggest that dynamical interactions, including the effects of the most massive stars and binary fraction, play a critical role in the formation and disruption of the layered structure in OCs.}


   \maketitle

\section{Introduction}

Open clusters (OCs) are gravitationally bound groups of stars formed after the collapse of giant molecular clouds. Their member stars typically share similar ages and chemical compositions. This makes OCs important objects for the study of star formation and evolution \citep{lada03, portegies10}. The ages of OCs range from a few million to nearly a billion years \citep{friel95}, and they are widely distributed throughout the Galactic disk. By studying the kinematics and spatial distribution of OCs, one can gain valuable insights into the Galactic gravitational potential and external perturbations \citep{soubiran18}. Moreover, the information provided by OCs can not only deepen our understanding of the dynamical and chemical evolution of star formation \citep{cantat-gaudin20}, but it also serves as a foundation for studying stellar mass distributions and Galactic structure \citep{adamo20}.

Open clusters were generally viewed as structures with high-density cores and low-density outer regions \citep{artyukhina64}, with the surrounding extended halo potentially containing a large number of member stars \citep{nilakshi02}. \citet{chen04} utilized Two Micron All Sky Survey data to classify clusters into central cores and outer regions, finding that the central region remains circular, while the outer region becomes increasingly elliptical. The differentiation of the core and outer regions is mainly due to different physical mechanisms in different regions. The structure of the core region is influenced by stellar evolution, two-body relaxation, and gravitational interactions \citep{vesperini09, portegies10}, while the outer region is more susceptible to external perturbations and Galactic tidal fields \citep{spitzer58, portegies10}.

Since 2018, high-precision astrometric data from the \textsl{Gaia} satellite \citep{gaia16} has fundamentally changed our understanding of OC morphology from an observational perspective. Covering approximately 1.8 billion sources brighter than 21 mag in the $G$ band \citep{gaia23}, the third data release (DR3) of \textsl{Gaia} marked a significant advancement. This dataset provides high-precision photometric measurements in three bands ($G$, $G_{BP}$, and $G_{RP}$) as well as astrometric solutions for 1.47 billion sources with five- or six-parameter solutions, thus improving parallax precision by 30\% compared to \textsl{Gaia} DR2 and nearly doubling the accuracy of proper motion measurements. Additionally, \textsl{Gaia} DR3 includes new radial velocity (RV) measurements for over 33 million stars \citep{katz23}, with median RV precisions of 1.3 km/s at GRVS = 12 mag and 6.4 km/s at GRVS = 14 mag. Furthermore, it provides atmospheric parameters ($T_{eff}$, log $g$, and [M/H]) for approximately 470 million sources \citep{fouesneau2023}. These advancements significantly improve the identification of cluster members and the precision of cluster structural analyses.

Recent studies utilizing \textsl{Gaia} data not only confirmed the existence of tidal tails but also revealed complex extended structures within OCs, including halos, tidal tails, and filamentary substructures \citep{roser19, zhang20, meingast21}. \citet{tarricq22} found that the overall size of OCs increases slightly with age, while the proportion of stars in the halo decreases. \citet{zhong22} introduced an improved model that utilizes \textsl{Gaia} EDR3 data to separately describe the core and halo components, further enhancing understanding of OC structure. 

Three-dimensional analysis is particularly important in the study of OC morphology. \citet{pang22} performed a statistical analysis of OCs within 500 pc using \textsl{Gaia} EDR3 data, finding that younger OCs exhibit elongated filament-like substructures, while older OCs display tidal-tail-like substructures. Meanwhile, \cite{hu23} explored the 3D spatial stratification of OCs and proposed a rose map structure, discovering that some clusters exhibit a layered structure in the 3D space.

Regarding dynamical evolution, $N$-body simulations are an important tool for studying this aspect of OCs \citep{aarseth03, wang15}, accurately simulated gravitational interactions between stars and external perturbations in order to reproduce changes in cluster morphology. \citet{chumak05} used $N$-body simulations to study the tidal tail structure of the Hyades, while \citet{jerabkova21} and \citet{boffin22} used {\footnotesize PeTar} to find that the tidal tails of the Hyades and NGC 752 are longer than observed. \cite{wang21} further utilized $N$-body simulations to investigate the impact of massive OB stars on the tidal evolution of young OCs, and their results indicate that the most massive star significantly influences the dynamical evolution of OCs.

In this work, we perform $N$-body dynamical simulations of nearby OCs within the Galactic potential to investigate the mechanisms behind the observed layered structure in 3D space. The paper is organized as follows. In Section \ref{section:2}, we provide an overview of the OC dataset used in this work and describe the methodology for analyzing the 3D stratification of OCs, including the $N$-body simulation tool and initial parameters. In Section \ref{section:3}, we present an analysis of the results from both the simulations and observations, and this is followed by a discussion. Finally, we summarize this work in Section \ref{section:4}.

\section{Data and 3D morphology}
\label{section:2}

The third data release of the \textsl{Gaia} survey \citep{gaia23} provides high-precision photometric and astrometric data. These data have substantially improved the accuracy and coverage of OC censuses. \citet{qin23} utilized \textsl{Gaia} DR3 data in combination with {\footnotesize pyUPMASK} and {\footnotesize HDBSCAN} clustering algorithms to publish a catalog of 324 OCs in the solar neighborhood within 500 pc. 

\citet{qin23} extended the search radius to identify members of OCs and reported 101 new OCs, resulting in a more complete catalog. The completeness of the catalog in terms of member stars enables the exploration of the 3D morphology in a wider region. Thus, we adopted the OC sample catalog published by \citet{qin23}. To ensure a robust analysis, we selected clusters with at least 50 member stars (\( N \geq 50\)) to be part of our sample. Ultimately, 279 OCs were selected for further analysis.

\subsection{3D layered structure}

\citet{hu23} introduced the rose map method for investigating the 3D layered structure of OCs. The rose map, characterized by its petal-like appearance, consists of multiple sectors with equal angular divisions but varying radial lengths. This method is also being applied in other areas, such as visualizing the distribution of velocity vectors in astronomical datasets \citep{meingast21}. Notably, it provides a clearer representation of spatial distributions, especially when analyzing complex 3D structures.

The process of creating the map involves two main steps. First, one uses the Python package \text{Astropy} \citep{astropy13,astropy18} to compute the galactocentric Cartesian coordinates (X, Y, Z) for each member star in the sample cluster. The transformation from International Celestial Reference System coordinates $(\alpha, \delta, d)$ to galactocentric Cartesian coordinates is defined as
\begin{equation}
\begin{aligned}
X &= R_{\odot} - d \cos(\delta) \cos(\alpha), \\
Y &= - d \cos(\delta) \sin(\alpha), \\
Z &= d \sin(\delta) + Z_{\odot}.
\end{aligned}
\end{equation}
where $(\alpha, \delta, d)$ denote the right ascension, declination, and distance of a member star, respectively. We adopted the Galactic center coordinates from \citet{reid04}, ${\text{Galactic center}} = (266^\circ.4051, -28^\circ.936175)$. The galactocentric distance of the Sun is 8.0 kpc, and the height of the Sun above the Galactic midplane is 15.0 pc. The peculiar motion of the Sun was adopted as (U, V, W) = (10.0, 235.0, 7.0) km/s. Since observational uncertainties may lead to elongation effects along the line of sight \citep{smith96, luri18, zhang20}, we avoided computing the distance directly as $d = 1000 / \varpi$ ($\varpi$ is the parallax from Gaia). Instead, we applied a Bayesian approach for parallax inversion \citep{bailerjones15, carrera19}. We modeled the prior distribution as an exponentially decreasing volume density function that incorporates cluster membership probabilities \citep{qin23}. The posterior probability distribution was then derived from the observed parallax and its uncertainty, with the posterior mean taken as the optimal distance estimate for each member star.

\begin{figure*}[h!]
\centering
\includegraphics[width=3in]{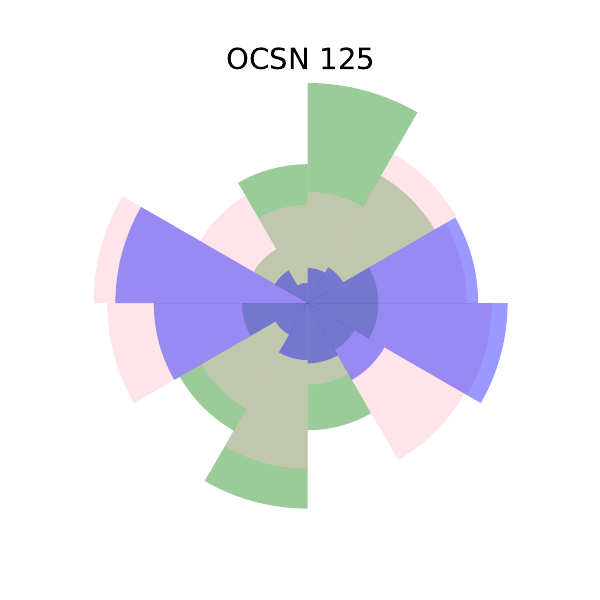} 
\includegraphics[width=3in]{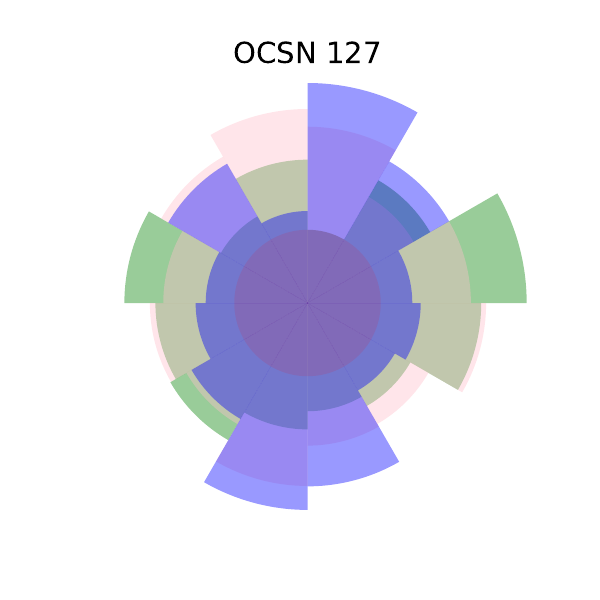}
\caption{Rose map of OCSN 125 and OCSN 127 obtained by superposition in 3D space, with the X-Y plane (green), X-Z plane (pink), and Y-Z plane (blue). The left subplot exhibits the absence of a layered structure, and the right subplot shows a layered structure.}
\label{fig1}
\end{figure*}

After obtaining the spatial coordinates of the cluster members, we applied the rose map method to analyze the three-dimensional spatial structure of the cluster. This method projects the spatial distribution of cluster members onto the X-Y, X-Z, and Y-Z planes, forming multiple sectors with different radii. Each plane is evenly divided into 12 equal-angle sectors, which allowed us to visualize the distribution characteristics of cluster members in different spatial regions. The radius of each sector is determined by two key parameters: the number of cluster members within the sector and the median distance of these members from the cluster center. By normalizing these two parameters, we obtained the radius value for each sector, thus quantifying the density and distribution of cluster members in different regions. When overlaying the rose map from multiple projection planes, the presence of a well-defined circular core suggests that the cluster exhibits a layered structure. Conversely, the absence of such a circular core indicates that the cluster lacks the layered structure. We defined $R$ as the radius characterizing the layered structure within each cluster, with $R = 0$ indicating a non-existent layered structure. Using the rose map method, we superimposed the OC samples in 3D space to assess their layered structure. The clusters with layered structures should have overlapping areas in the center of their rose maps. The rose maps for the OCSN 125 and OCSN 127 are shown in Figure \ref{fig1}. As is visible from the figure, OCSN 127 displays a layered structure ($R \neq 0$), while OCSN 125 lacks overlapping areas, indicating no layered structure ($R = 0$).

\begin{figure}[h!]
    \centering
    \includegraphics[width=1\linewidth]{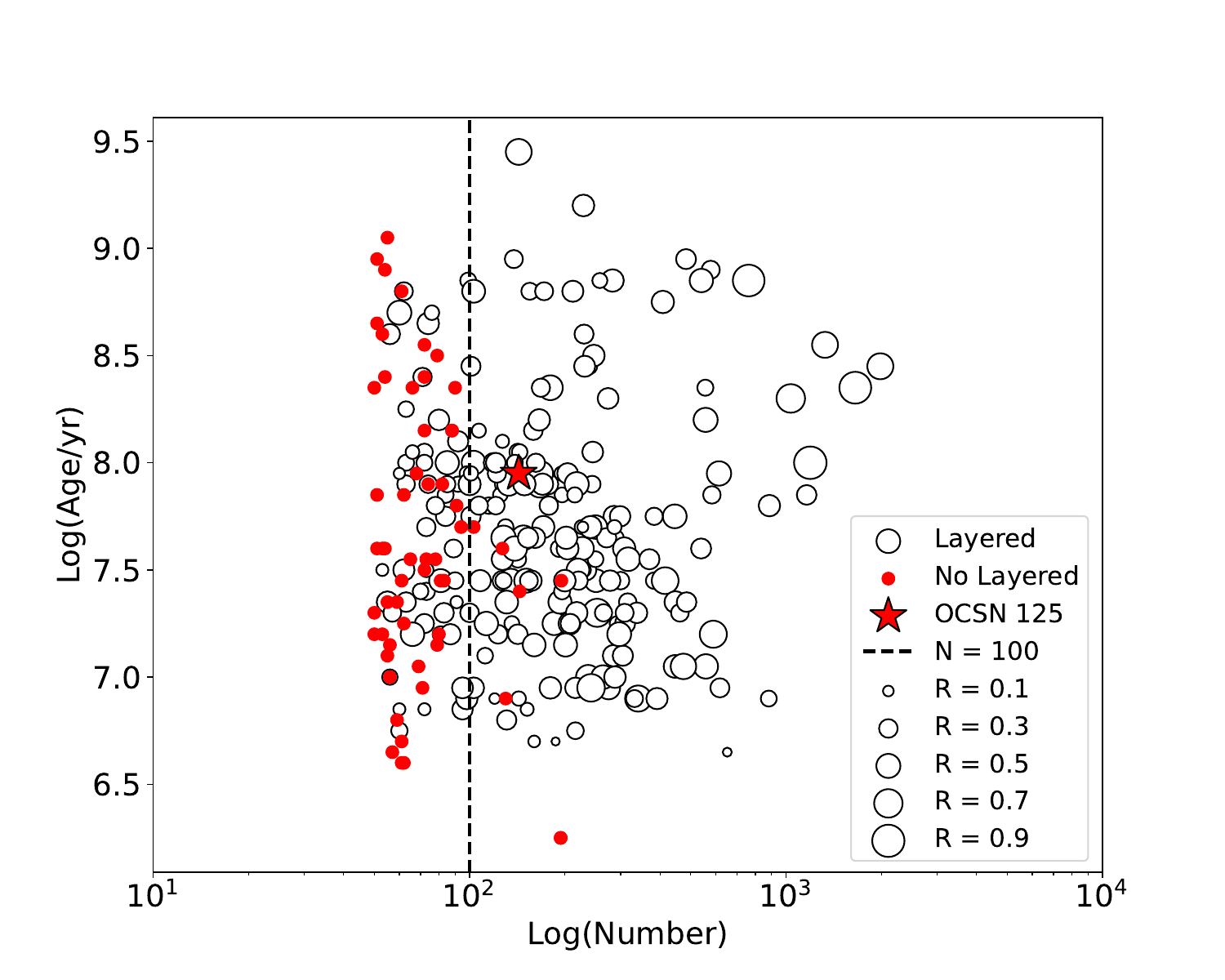}
    \caption{Number of member stars in the 279 OC samples shown in relation to cluster age and the presence of a layered structure. The red filled circles represent clusters without a 3D layered structure, while empty circles indicate those with such a structure. The size of the circles corresponds to the radius of the layered structure region. The red star denotes the selected cluster OCSN 125.}
    \label{fig2}
\end{figure}

We carried out the rose map analysis on all OC samples to investigate the formation pattern of the layered structure. Statistical analysis was performed on 279 OCs; the results are presented in Figure \ref{fig2}. Analysis revealed a significant correlation between the layered structure of clusters and the total number of member stars. We found that clusters lacking a layered structure typically have fewer than 100 member stars, and only seven clusters with more than 100 member stars lacked such structures. In contrast, a layered structure was observed across clusters of various ages, and no significant differences were shown. We selected the OCSN 125 for further investigation into the potential reasons behind this phenomenon, as this object has a relatively large number of member stars and an intermediate age.

\subsection{$N$-body simulation}

To investigate the underlying causes of the layered structure in the OCs, we employed $N$-body simulations to explore the dynamical evolution of the cluster OCSN 125. $N$-body simulation is a fundamental technique in computational astrophysics, and it is used to study the dynamical evolution of systems with multiple interacting bodies, such as clusters, galaxies, and large-scale cosmic structures \citep{aarseth03, hurley05}.

We utilized {\footnotesize PeTar},\footnote{https://github.com/lwang-astro/PeTar} a high-performance $N$-body simulation tool designed for simulating the dynamical evolution of clusters \citep{wang20a}. Built on the architecture of the Framework for Developing Particle Simulators ({\footnotesize FDPS}), {\footnotesize PeTar} takes advantage of multi-core parallel computing, thus enabling excellent performance on modern computing systems \citep{iwasawa16, namekata18, iwasawa20}. Key features of {\footnotesize PeTar} include advanced algorithms for binary star dynamics, such as regularization techniques \citep{wang20b}, and integration with well-established stellar and binary evolution codes, {\footnotesize SSE} and {\footnotesize BSE} \citep{hurley00, hurley02, banerjee20}. {\footnotesize PeTar} also employs the initial mass function (IMF) to define the initial mass distribution of stars \citep{kroupa01}, with masses ranging from $0.08$\,M$_\odot$ upwards. The gravitational potential of the Milky Way was considered using the {\footnotesize GALPY} code with the {\footnotesize MWPotential2014} Galactic potential model \citep{bovy15}. 

\begin{figure}[!ht]
    \centering
    \includegraphics[width=1\linewidth]{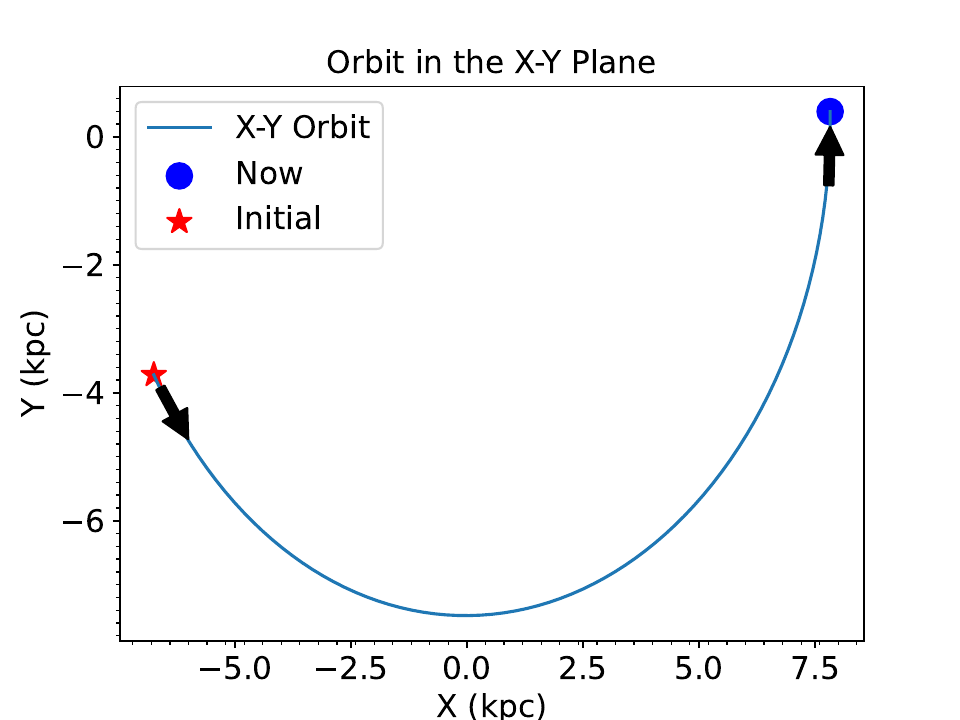}
    \caption{Past motion of OCSN 125 motion through the Milky Way using the {\footnotesize GALPY}. The plot shows the orbit of OCSN 125 projected in the X-Y plane. The red star indicates the cluster's starting position, and the blue point represents its current location. The arrows show the direction of motion of OCSN 125 in the X-Y plane.}
    \label{fig3}
\end{figure}

We used {\footnotesize GALPY} to integrate the orbital trajectory of OCSN 125, and we determined the birthplace and initial velocity. The actual trajectory of the OCSN 125 cluster within the Milky Way over the past 89 Myr is illustrated in Figure \ref{fig3}. The initial coordinates and velocities at galactocentric Cartesian coordinates are as follows: 

[X,Y,Z] = -7.394, -4.111, 0.161 kpc;

[Vx,Vy,Vz] = 109.381, -176.574, -2.459 km/s.

\noindent Under the current {\footnotesize GALPY} code, the position and coordinates of the Sun are as given below:

[X,Y,Z] = -8.00, 0.0, 0.015 kpc;

[Vx,Vy,Vz] = 10.0, 235.0, 7.0 km/s.

We determined the mass of the observed member stars using the color-magnitude diagram (CMD) of the cluster. The apparent magnitude limit in the $G$ band was set at 18, as \textsl{Gaia} data are highly complete within this range. The CMD of the cluster OCSN 125 with the best-fit \citep{marigo17} isochrone is shown in Figure \ref{fig4}. 

According to \citet{qin23}, OCSN 125 has equatorial coordinates  ($\alpha$, $\delta$) = ($233^\circ.57$, $-29^\circ.70$). Its proper motion is ($\mu_{\alpha}^*$, $\mu_{\delta}$) = ($-6.84$, $-5.02$) mas/yr. The parallax is $ \varpi$ = $2.60$ mas, with a distance modulus of $DM = 8.29$ mag. The cluster contains $N$ = 143 member stars, with an angular size of half-number radius $r_h$ = $3.93$ deg. The radial velocity is $RV$ = $-26.38$ km/s. The age of the cluster is 89 Myr, with a metallicity value of $Z$ = 0.0152 and an extinction of $E(B-V) = 0.17$ mag. To obtain the mass of each star, we determined the nearest point on the best-fit isochrone and assigned the corresponding mass as the star mass. We then determined the initial total mass of the cluster by applying the multiple power-law mass function provided by \citet{kroupa01}. This method is similar to those described by \citet{snider09}, \citet{maurya23}, and \citet{elsanhoury25}. According to this power-law, the number of stars between the mass interval $m_1$ and $m_2$ could be given by the following relation:

\begin{figure}[!ht]
    \centering
    \includegraphics[width=1\linewidth]{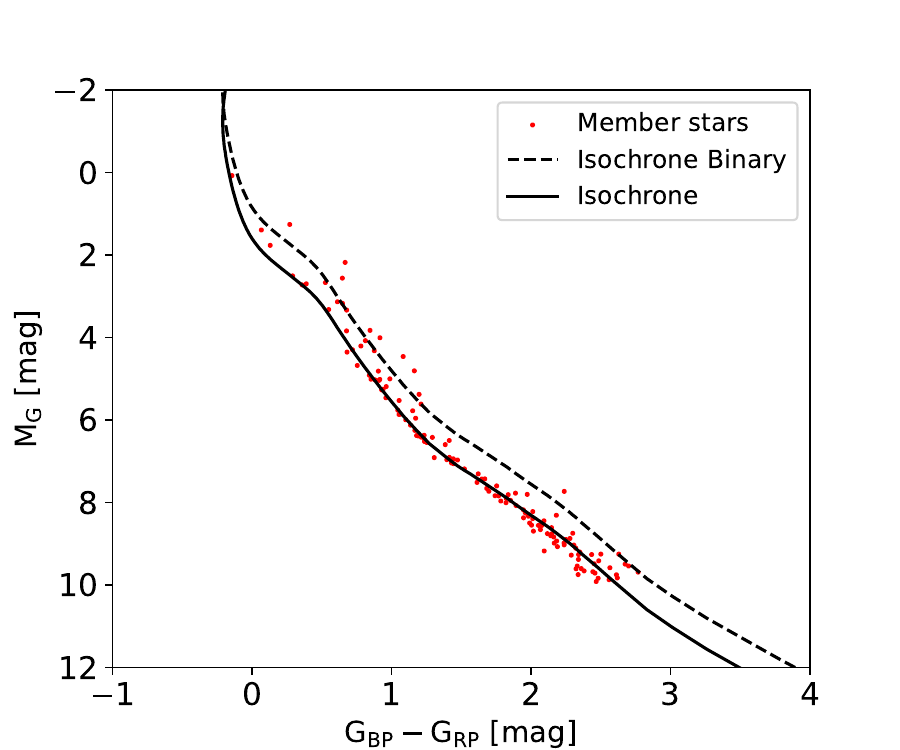}
    \caption{Plot for the CMD of the cluster OCSN 125. The black solid line and dotted line represent the best-fit isochrones corresponding to the single stars and the equal mass-ratio binary stars, respectively. The red points mark the member stars of OCSN 125.}
    \label{fig4}
\end{figure}

\begin{equation}
N = {A}_{i}\times \int_{{m}_{1}}^{{m}_{2}}{{M}^{-\alpha }}dM,
\end{equation}where $N$ represents the number of stars within the specified mass interval, $m_1$ and $m_2$ are the lower and upper mass limits, and $\alpha$ is the exponent of the power-law in the IMF \citep{kroupa01}. From this equation, we calculated the normalization constant $A_i$ for the mass interval m $\geq$ 1.0\,M$_\odot$.

\begin{equation}
{M}_{tot} = \int_{0}^{N}{M}dN = {A}_{i}\times \int_{{m}_{1}}^{{m}_{2}}{{M}^{1-\alpha }}dM.
\end{equation} 

Using the above equation, we calculated the total mass of stars in the mass interval $m$ $\geq$ $1.0$\,M$_\odot$. The normalization constant $A$$_{i}$ values for the mass intervals $0.08-0.5$\,M$_\odot$ and $0.5-1$\,M$_\odot$ were calculated by utilizing the relationships among the normalized constants for these mass ranges with that of the $m$ $\geq$ $1.0$\,M$_\odot$ mass interval given by \citet{maschberger13}. We then summed the masses corresponding to the mass intervals $0.08-0.5$\,M$_\odot$, $0.5-1$\,M$_\odot$, and $1-100$ M$_\odot$ to obtain the total mass of the cluster. Following this process, we estimated the total initial mass to be $M_{tot} = 224 \pm 138$\,M$_\odot$ for the cluster OCSN 125.

In this work, our simulations start at the gas-free phase, leading to systems that are composed solely of stellar particles. We utilized the latest version of the particle generation tool {\footnotesize MCLUSTER}\footnote{https://github.com/lwang-astro/mcluster} to create the initial conditions of our simulated clusters \citep{kupper11}. The structures of OCs are influenced by various factors, such as the evolution of stars within molecular clouds, stellar interactions, and the external gravitational potential of the Milky Way. However, the Plummer model provides an effective method for characterizing the structure of a gas-free cluster \citep{plummer11, roser11, roser19}. Therefore, we adopted the Plummer density profile to randomly generate the initial 3D positions and velocities of the simulated stars \citep{aarseth74}. The initial parameters of the cluster are given in Table \ref{table1}. The mass distribution of the simulated cluster follows the \citet{kroupa01} mass function in the mass range of $0.08-100$\,M$_\odot$. The initial half-mass radius of 15 pc, including all star components in three-dimensional space, is the most consistent with the spatial distribution of the observed data. The cluster metallicity and the initial number of the member stars were taken as $Z$ = 0.0152 and $N$ = 400, respectively. We considered the evolution time to be 89 Myr and three binary fractions: 0, 50\%, and 100\%. The initial number of members was set to 400 to ensure that the total initial mass of the cluster remains controlled, thereby preventing the presence of extremely massive stars from dominating the cluster's mass and potentially affecting the overall number of members. We repeated the simulation 100 times for each binary fraction.

\begin{table}[h]
    \centering 
    \caption{Initial conditions used for the {\footnotesize PeTar} simulations.}
    \begin{tabular}{cc}
    \hline
    \hline  
    Parameters      & Initial conditions       \\ \hline
    Number of member star  & 400                      \\
    Half-mass radius  & 15 pc                      \\
    Binary fraction & 0                      \\
    Binary fraction & 50\%                     \\
    Binary fraction & 100\%                    \\
    Metallicity     & 0.0152                     \\
    {[}X,Y,Z{]}     & -7.394, -4.111, 0.161 kpc \\
    {[}Vx,Vy,Vz{]}  & 109.381, -176.574, -2.459 km/s   \\ \hline
    \end{tabular}
    \label{table1}
\end{table}

\section{Result and discussion}
\label{section:3}

\subsection{Comparison of simulation and observation}
\label{section:3.1}

\begin{figure*}[htbp]
    \centering
    \includegraphics[width=0.9\textwidth]{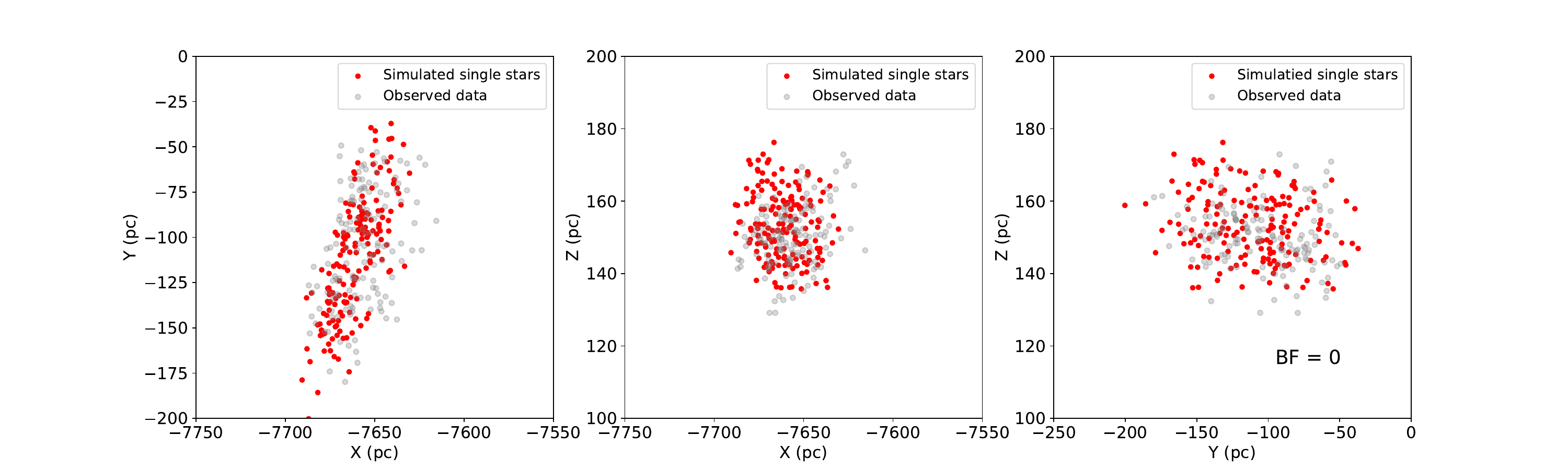} 
    \includegraphics[width=0.9\textwidth]{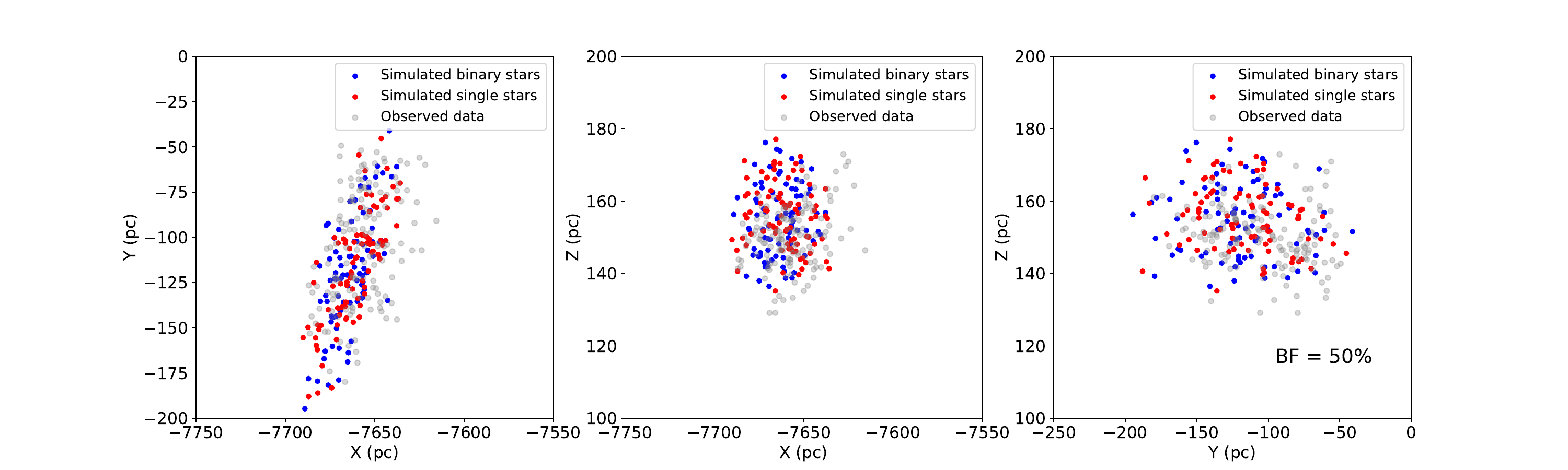}
    \includegraphics[width=0.9\textwidth]{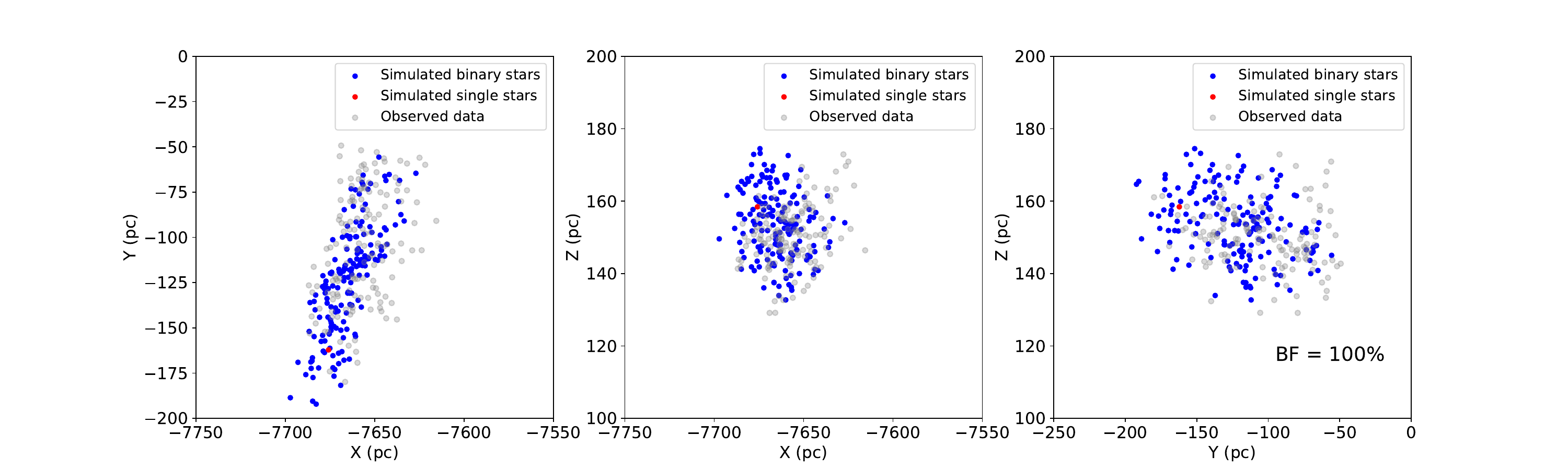}
    \caption{Comparison of the spatial distribution between simulated data with varying initial binary fractions and observed data for OCSN 125. From top to bottom, the rows represent simulations with initial binary fractions of 0, 50\%, and 100\%. From left to right, the columns display the X-Y, X-Z, and Y-Z coordinate projections. Red points indicate binary star systems, blue points represent single stars, and gray points correspond to the observed data.}
    \label{fig5}
\end{figure*}

To assess the reliability of our simulations, we compared the spatial distribution of simulated cluster populations in the OC with varying initial binary fractions to the observed member stars. The comparison results of the spatial distribution are shown in Figure \ref{fig5}. The lower mass limit of the simulated member was taken as $0.3$\,M$_\odot$ to cope with the lower mass limit of the observed member stars according to the completeness of the \textit{Gaia} DR3 data. The spatial distribution of the simulated populations closely resembles that of the observed distribution, and the number of internal members is similar. This indicates that the simulations not only reproduce the spatial characteristics of clusters with different binary fractions but also effectively represent the underlying dynamical processes and distribution mechanisms in real clusters. We used the {\footnotesize EMCEE} method \citep{foreman-mackey13}, implemented in {\footnotesize PYTHON}, to analyze correlations in the simulated cluster with different binary fractions. Additionally, we examined the correlation values between parameters within the simulated clusters derived from {\footnotesize EMCEE} for different binary fractions. To further ensure the accuracy and reliability of the correlation values, we calculated p-values from Pearson correlation tests \citep{pearson01}.

\begin{figure*}[htbp]
    \centering
    \includegraphics[width=0.9\linewidth]{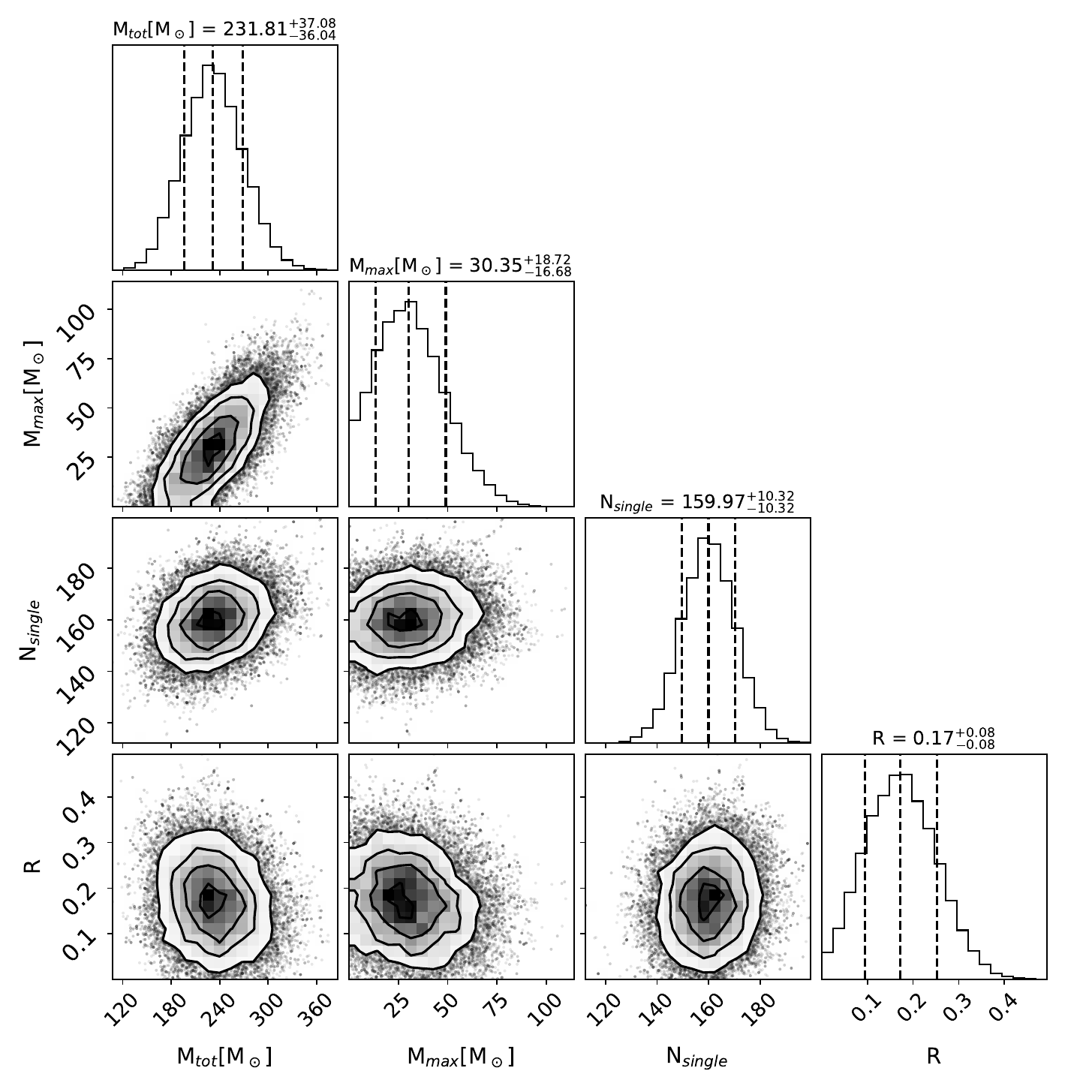}
    \caption{{\footnotesize EMCEE} parameter sampling of the simulated clusters with an initial binary fraction of 0\% at 89 Myr. The first column shows the total initial mass of the simulated cluster, the second column depicts the mass of the most massive star in the simulation, and the last column represents the radius of the layered structure. The remaining columns represent the number of single stars, and the vertical dashed black lines indicate the one standard deviation spread at the 16th, 50th and 84th percentiles.}
    \label{fig6}
\end{figure*}

The {\footnotesize EMCEE} correlation analysis results for different binary fractions are shown in Figures \ref{fig6}, \ref{fig7}, and \ref{fig8}. In each simulation, we selected the most massive star in the cluster and observed a significant negative correlation between the layered structure radius and the mass of the most massive star. This relationship became more pronounced with the increasing binary fraction. The correlation coefficients and Pearson p-values for each parameter pair are summarized in Table \ref{table:combined}. For clusters composed entirely of single stars, the correlation between the layered structure radius and the mass of the most massive star is weak, with a correlation coefficient of -0.135 and a p-value of 0.205. This suggests that in such clusters, although the most massive star may cause a slight decrease in the layered structure radius, the effect lacks statistical significance. When the initial binary fraction is increased to 50\%, the negative correlation between the mass of the most massive star and the layered structure radius becomes more statistically significant, with a correlation coefficient of -0.278 and a p-value of 0.009, indicating that the presence of binaries enhances this relationship. In the case of a binary fraction of 100\%, the negative correlation is further strengthened, with a correlation coefficient of -0.402 and a p-value of 0.00002. This trend suggests that, in clusters composed entirely of binary stars, the mass of the most massive star significantly reduces the layered structure radius. These findings reveal two key features: (1) The most massive star in the cluster may constrain the layered structure radius, with the layered structure becoming less pronounced as its mass increases, and (2) this negative correlation is strengthened with an increasing binary fraction. This suggests that a higher binary fraction and the higher mass of the most massive star in the cluster play important roles in shaping the spatial distribution and evolution of clusters.

\begin{table*}[!ht]
    \centering
    \caption{Correlation between the layered structure radius and various parameters along with a comparison of p-values under different initial binary fractions from simulations.}
    \setlength{\tabcolsep}{5pt}
    \begin{tabular}{lccc}
    \hline
    \hline
        Quantities compared & Initial binary fractions & Correlation coefficient & p-value \\ 
    \hline
        R vs. M$_{tot}$[M$_\odot$] & 0 & -0.151 & $0.123$  \\ 
        R vs. M$_{max}$[M$_\odot$] & 0 & -0.135  & $0.205$  \\ 
        R vs. N$_{single}$ & 0 & 0.093 & $0.271$ \\ 
    \hline
        R vs. M$_{tot}$[M$_\odot$] & 50\% & -0.247 & $0.011$   \\ 
        R vs. M$_{max}$[M$_\odot$] & 50\% & -0.278  & $0.009$  \\ 
        R vs. N$_{binary}$ & 50\% & 0.186 & $0.018$  \\ 
        R vs. N$_{single}$ & 50\% & 0.011 & $0.479$  \\ 
    \hline
        R vs. M$_{tot}$[M$_\odot$] & 100\% & -0.299 &  $0.0005$ \\ 
        R vs. M$_{max}$[M$_\odot$] & 100\% & -0.402 &  $0.00002$\\ 
        R vs. N$_{binary}$ & 100\% & -0.081 &  $0.637$ \\ 
        R vs. N$_{single}$ & 100\% &  0.053 &  $0.994$\\ 
    \hline
    \end{tabular}
    \label{table:combined}
\end{table*}

\subsection{The effect of massive stars}

According to the correlation study results in Table \ref{table:combined}, the most massive star in the simulated OCSN 125 can greatly influence the layered structure. Clusters with an initially very massive star may exhibit a reduced degree of spatial stratification. For clusters without a layered structure, the most massive star exceeds $8$\,M$_\odot$. Several factors associated with the most massive star may explain this interesting outcome.

Massive stars, which typically exceed $8$\,M$_\odot$, end their lives in supernova explosions \citep{woosley86a, woosley86b, smartt09}. The feedback activities of massive stars, including supernovae and stellar winds, can release large amounts of energy. These events eject the outer layers of the star into the interstellar medium, releasing vast amounts of energy and material. Massive stars can significantly influence cluster dynamics, particularly during the early stages when they are concentrated in the core \citep{langer12, gehrman19}, and supernova explosions cause mass loss within the cluster, thus altering the dynamic balance of the cluster. These explosions could impede the formation of spatial stratification by disrupting the internal structure of the cluster \citep{goodwin06, baumgardt07}. In addition to the disruptive impact of supernovae, strong stellar winds from massive stars lead to sustained mass loss, further altering the cluster's dynamics and spatial stratification \citep{portegies02}. Combined with supernova explosions, these processes accelerate the cluster's evolution. As massive stars lose a substantial portion of their mass early in their evolution, the gravitational binding of the cluster's core weakens, leading to a hotter and more dynamically active core. This heightened activity prevents core collapse, which in turn inhibits the development of pronounced spatial stratification \citep{lamers05}. 

\citet{mapelli13} and \citet{trani14} found that core collapse can be reversed by mass loss due to stellar winds and supernovae. Moreover, recent studies using $N$-body simulations have demonstrated that clusters with identical initial conditions can evolve along divergent paths \citep{mackey08, wang21}. \citet{wang21} employed the {\footnotesize PeTar} to simulate the tidal stream structure of the Hyades. Under unified simulation conditions, they found that differences in the initial mass of the massive stars led to significant variations in the evolved structure and dynamical state of the cluster. This finding aligns with the results of the present study, indicating that the presence of massive stars and the mass loss caused by stellar winds can significantly influence the structural and dynamical evolution of clusters. Such mass loss may accelerate the overall evolutionary process and impact the emergence of a layered structure in clusters.

In conclusion, our simulations suggest that the mass of the most massive star in the simulated OCSN 125 may play a critical role. It influences the degree of 3D spatial stratification within the cluster, especially during its early stages of development. As indicated in Table \ref{table:combined}, clusters with more massive stars initially appear to exhibit smaller layered structural radii. For instance, in clusters with a binary fraction of 100\%, the correlation coefficient reaches -0.402, which may suggest a constraining influence of the most massive star on spatial dynamics. Additionally, our analysis of Figures \ref{fig6}, \ref{fig7}, and \ref{fig8} revealed that clusters initially with massive stars could show reduced spatial stratification and indicate more uniform star distributions. These findings imply that the mass loss from supernova explosions and strong stellar winds can accelerate the evolution of OCSN 125 while potentially inhibiting the development of a distinct layered structure.

Due to feedback from massive stars that may end their lives as supernovae, the presence of such stars in OCSN 125 indicates strong internal feedback and ongoing dynamical evolution. To evaluate whether OCSN 125 has reached dynamical equilibrium, we calculated the half-mass relaxation time using the standard expression from \citet{spitzer87}:
\begin{equation}
T_{\text{rh}} \approx 0.138 \frac{N^{1/2} r_{\text{h}}^{3/2}}{m^{1/2} G^{1/2} \ln \lambda}
\end{equation}
where \( N \) is the number of stars, \( r_{\text{h}} \) is the half-mass radius, \( m \) is the average stellar mass, \( \lambda \) accounts for tidal effects, and \( G \) is the gravitational constant. We find that \( T_{\text{rh}} \approx 1636.88 \) Myr, which is much longer than the current age of the cluster (89 Myr), indicating that OCSN 125 remains in a dynamically non-relaxed state. We also estimated the escape velocity using the formula from \citet{maeda24}, \( V_{\text{esc}} = \sqrt{2GM/r_{\text{h}}} \), where \( M \) is the total cluster mass. The resulting escape velocity is 0.18 km/s, which is substantially lower than the observed velocity dispersion of \( 1.27 \pm 0.21 \) km/s. This discrepancy suggests that a significant fraction of stars are unbound, reinforcing the conclusion that OCSN 125 has not yet reached dynamical relaxation.

\subsection{The effect of binary fraction}

Our results presented in Section \ref{section:3.1} suggest that the initial binary fraction in the simulated OCSN 125 may play a significant role in shaping the 3D spatial structure. Analysis of Table \ref{table:combined} indicated that clusters with higher initial binary fractions tend to exhibit a more uniform stellar distribution, which could imply a reduction in spatial stratification. This phenomenon might arise from the dynamical influences that binary star systems exert, potentially leading to a more homogeneous energy distribution within the cluster.

Binary star systems may influence the evolution of star clusters through gravitational interactions and energy exchanges \citep{heggie75, hills75}. \citet{heggie03} described how these interactions contribute to a more uniform energy distribution within the cluster, as they reduce energy gradients between stars. Furthermore, the gravitational perturbations caused by binary systems might increase the randomness and chaotic nature of stellar motions within the cluster. This perturbative effect could result in a more randomized orbital distribution, potentially diminishing the likelihood of forming a layered structure. Additionally, the increased frequency of stellar collisions and energy exchanges associated with binary systems may further contribute to the overall dynamical uniformity of the cluster \citep{heggie03, binney11}.

In clusters with a higher fraction of binary systems, dynamical friction may play an important role in altering the structure of the cluster. Dynamical friction occurs when a member star moves through the cluster due to gravitational interactions. When massive objects, such as binary stars, move through the cluster, their gravitational interactions with surrounding lower mass stars may cause the binaries to lose energy \citep{heggie03}. As binary systems migrate inward, they may increase the stellar density in the core region, possibly disrupting any initial stratification. The enhanced central concentration of mass could create a more uniform distribution of stars, thus making it harder for distinct spatial layers to form. Higher binary fractions may amplify this effect, leading to reduced spatial stratification and a more isotropic stellar distribution \citep{spitzer87, portegies01, heggie03}.

The increased kinetic energy provided by binary systems through gravitational interactions and dynamical friction could potentially slow down the core collapse process \citep{gao91}. According to \citet{hurley05}, core collapse is not distinctly observed in binary-rich star clusters. \citet{portegies01} and \citet{fregeau03}, using N-body simulations found that higher initial binary fractions may benefit from the increased kinetic energy from binaries, potentially delaying core collapse of clusters and mitigating stratification, which could result in a more uniform cluster distribution. \citet{kremer19} mentioned that the core collapse can be halted by an energy source in the core, which is expected to arise from binaries.

In conclusion, our simulation results indicate that the proportion of binary star systems present during evolution may play a crucial role in influencing the degree of 3D spatial stratification in the simulated OCSN 125. Analysis of Figures \ref{fig6}, \ref{fig7} and \ref{fig8} revealed a trend: Clusters with higher initial binary fractions may exhibit diminished spatial stratification, potentially resulting in a more homogenous stellar distribution. These findings imply that the dynamical interactions and energy exchanges associated with binary systems may variably influence the evolutionary processes of clusters.

\section{Summary} 
\label{section:4}
In this paper, we have applied the rose map method to examine the layered structure of 279 OCs and found that the number of member stars in OCs has a certain influence on the layered structure of these clusters in 3D space. We observed that OCs with more than 100 member stars typically display a layered structure, with only a few exceptions lacking this feature. To investigate the factors behind this spatial feature, we focused on OCSN 125 and created $N$-body simulated cluster populations with varying initial binary fractions. After comparing the simulation results with the observed cluster populations, we find that the 3D stratification in OCSN 125 is likely influenced by several factors, such as the most massive star and initial binary fractions.

Our results suggest that the most massive star may play a significant role in reducing the degree of spatial stratification within OCSN 125. This potential reduction could be attributed to mass loss caused by supernova explosions and stellar winds, which disrupt the gravitational binding of the cluster and prevent the formation of a spatial layered structure.

The initial binary fractions may also significantly influence the presence of spatial stratification. Clusters with higher binary fractions appear to show less pronounced layering, likely due to the equipartition of energy, dynamical friction, and perturbations associated with binary systems. These factors can inhibit core collapse and contribute to the homogenization of the internal dynamics of the cluster, leading to a reduction in layered structure.

Our work provides valuable insights into the formation mechanism of the layered structure, revealing a complex interplay among massive stars, binary systems, and cluster evolution. It suggests that considering initial conditions and binary fractions in simulations may influence the spatial distribution of stars within the cluster, potentially affecting the development of a layered structure. However, this study focuses only on the simulation of OCSN 125, which limits the generalizability of our findings. Future work will aim to expand the sample by including additional clusters, thus allowing for a more comprehensive analysis. By incorporating a wider range of initial conditions and clusters, we hope to further validate and refine our understanding of cluster dynamics.

\begin{acknowledgements}
The authors thank the reviewer for the very helpful comments and suggestions. K.L. thanks Long Wang for helpful discussions on the {\footnotesize PeTar} tool. The authors acknowledge the Chinese Academy of Sciences (CAS) "Light of West China" Program, No. 2022-XBQNXZ-013, the National Key R\&D program of China for Intergovernmental Scientific and Technological Innovation Cooperation Project under No.2022YFE0126200, the Tianshan Talent Training  Program through the grant 2023TSYCCX0101, the  National Natural Science Foundation of China NSFC 12433007, the Natural Science Foundation of Xinjiang Uygur Autonomous Region, No.2022D01E86 and No.2024D01B89, and the Tianchi Talent project for providing additional support. This work has made use of the Gaia DR3 from the European Space Agency (ESA) space mission \textsl{Gaia} (\url{https://www.cosmos.esa.int/gaia}), processed by the \textsl{Gaia} Data Processing and Analysis Consortium (DPAC, \url{https://www.cosmos.esa.int/web/gaia/dpac/consortium}). Funding for the DPAC has been provided by national institutions; in particular, the institutions participating in the \textsl{Gaia} Multilateral Agreement. Software: {Astropy \citep{astropy13,astropy18}, {\footnotesize PeTar} \citep{wang20a}, {\footnotesize MCLUSTER} \citep{kupper11}, Matplotlib \citep{baumgardt07}, NumPy \citep{harris20}, {\footnotesize GALPY} \citep{bovy15}}
\end{acknowledgements}

\begin{appendix}
\onecolumn
\section{Layered structure correlation analysis figures}

\begin{figure*}[htbp]
    \centering
    \includegraphics[width=0.56\linewidth]{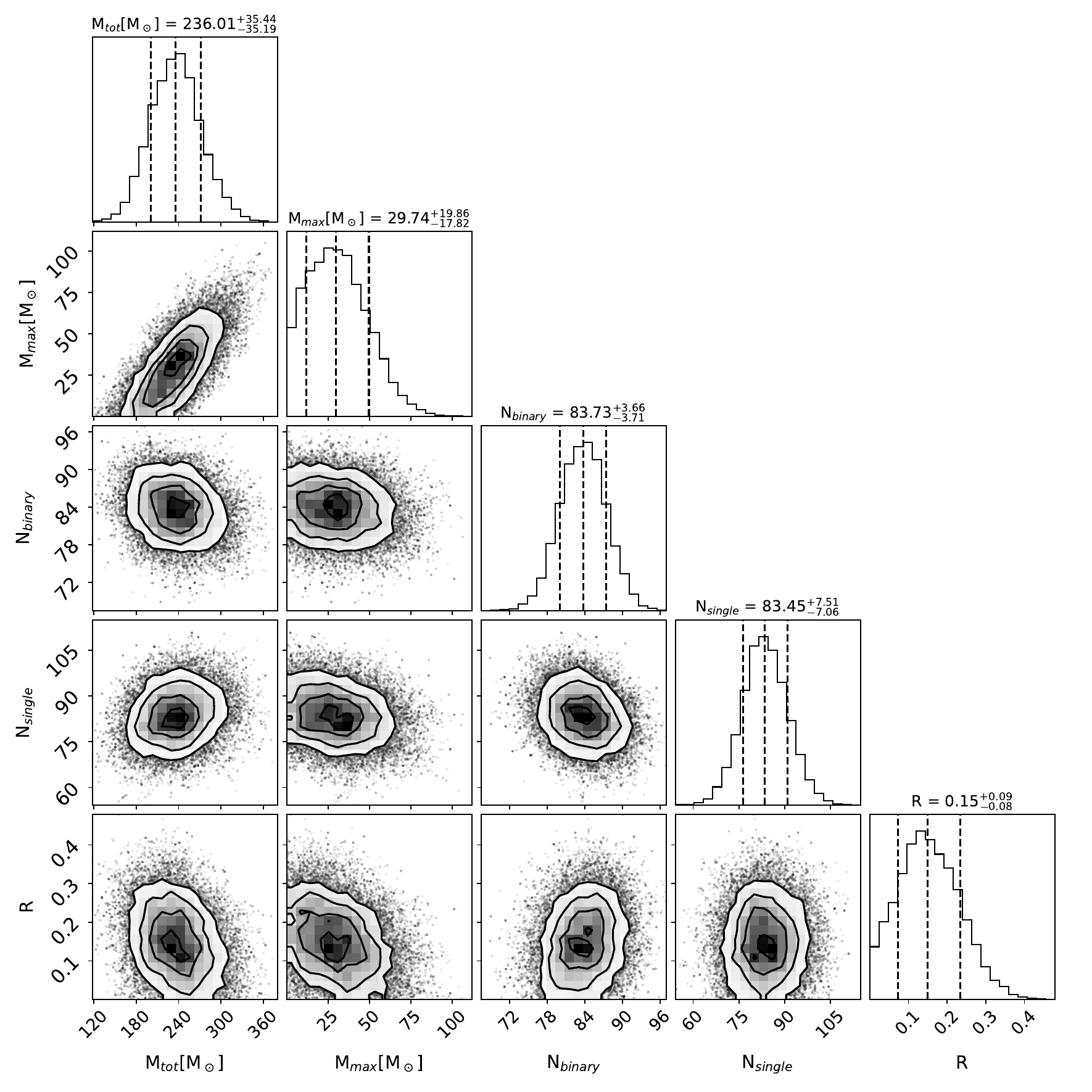}
    \caption{The same as Fig \ref{fig6}, but for the simulation with an initial binary fraction of 50\%.}
    \label{fig7}
\end{figure*}

\begin{figure*}[htbp]
    \centering
    \includegraphics[width=0.56\linewidth]{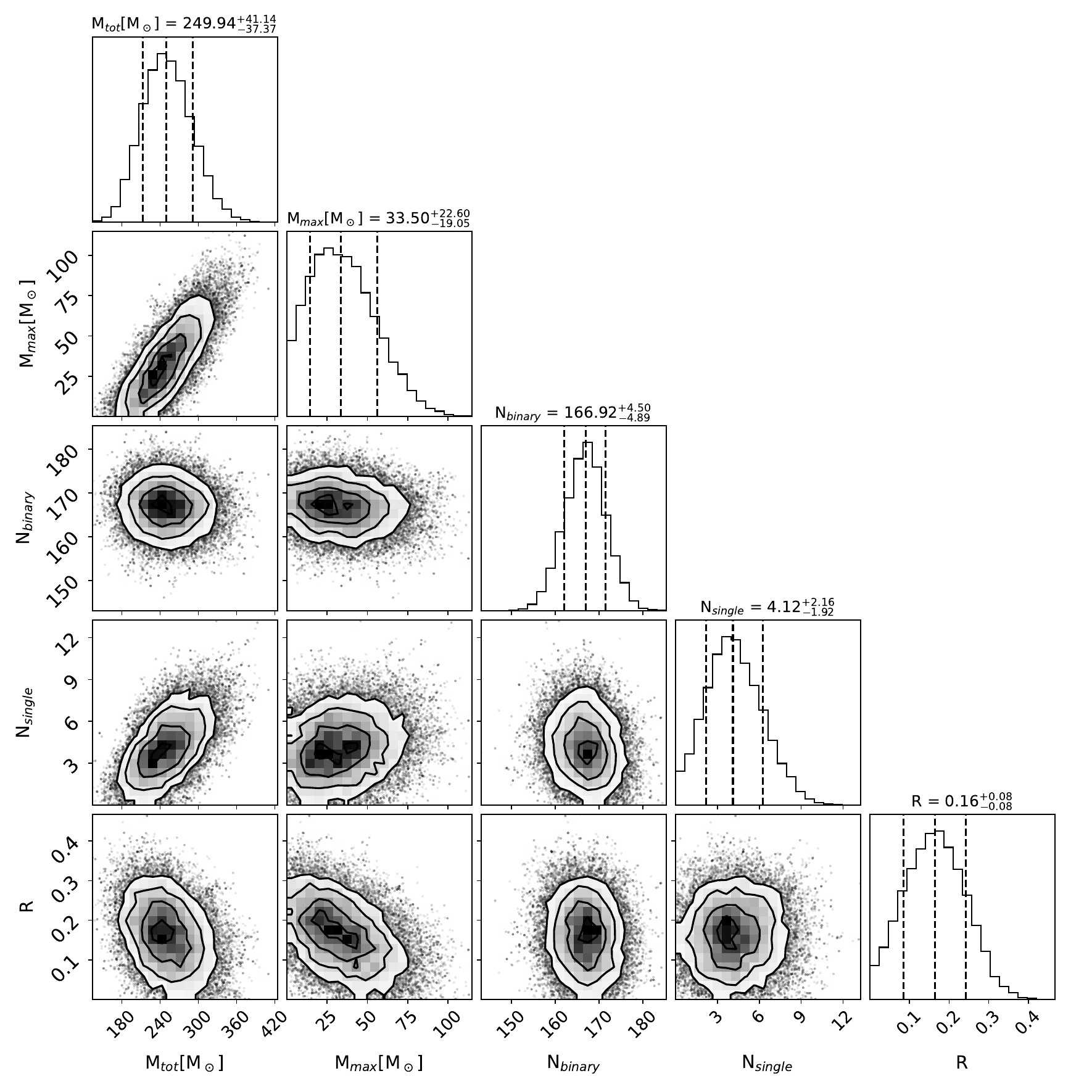}
    \caption{The same as Fig \ref{fig6}, but for the simulation with an initial binary fraction of 100\%.}
    \label{fig8}
\end{figure*}

\end{appendix}

\FloatBarrier
\clearpage
\end{CJK*}
\end{document}